\documentclass[10pt,two column,pre]{revtex4-1}
\usepackage{xcolor}
\usepackage[utf8]{inputenc}
\usepackage{fancyhdr} 

\usepackage{amsmath}
\usepackage{amsfonts}
\usepackage{amssymb}
\usepackage{graphicx,hyperref}
\usepackage{amssymb}
\usepackage{mathptmx}

\newcommand{\be}{\begin {equation}}
\newcommand{\ee}{\end {equation}}

\newcommand{\atitle}[1]{\begin{center}{\sc\Large #1}\\[0.5cm]\end{cgenter}}

\newcommand{\mb}{\mathbf}
\hypersetup{
    colorlinks,    citecolor=blue,    filecolor=black,    linkcolor=blue,    urlcolor=blue
}

\newcommand{\Amol}[1]{{\textcolor[HTML]{0000CD}{#1}}}

\begin{document}
 
\title{\Large{Dispersion    of   the    laser   pulse    through   propagation    in   underdense    plasmas}}
\author{Shivani  Choudhary}  \email{shivani.choudhary@pilani.bits-pilani.ac.in}  \author{Amol  R.  Holkundkar}
\email{amol.holkundkar@pilani.bits-pilani.ac.in}

\affiliation{Department of  Physics, Birla Institute  of Technology and  Science - Pilani,  Rajasthan, 333031,
India}

\begin{abstract} The propagation of the laser pulses in the underdense plasma is a very crucial aspect of
laser-plasma interaction process. In this work, we explored the two regimes of laser propagation in plasma, one
with $a_0 < 1$ and other with $a_0 \gtrsim 10$. For $a_0<1$ case, we used a cold relativistic fluid model,
wherein apart from immobile ions no further approximations are made. The effect of the laser pulse amplitude,
pulse duration, and plasma density is studied using the fluid model and compared with the expected scaling laws
and also with the PIC simulations. The agreement between the fluid model and the PIC simulations are found to
be excellent. Furthermore, for $a_0 \gtrsim 10$ case, we used the PIC simulations alone. The delicate interplay
between the conversion from the electromagnetic field energy to the longitudinal electrostatic fields results
in the dispersion and so the red-shift of the pump laser pulse. We also studied the interaction of the
dispersed pulse (after the propagation in underdense plasma) with the sub-wavelength two-layer composite
target. The ions from the thin, low-density second layer are found to be efficiently accelerated to $\sim 70$
MeV, which is not found to be the case without dispersion.
\end{abstract}

\date{\today}

\maketitle

\section{Introduction}

Since  the last  couple of  decades  we  are  witnessing the  rapid  technological  advancement  in the  field  of
high  power  lasers,  promising  number  of  applications  in  both  applied  and  fundamental  sciences.  The
table-top  setup  for  the ion  and  electron  accelerations  to  relativistic  energies  is a  result  of  the
technological breakthrough  in the field of  high power lasers. The  idea of the laser  wakefield acceleration
as  demonstrated   in  Ref.  \cite{Mangles2004_Nature,Faure2004_Nature,Geddes2004_Nature}  really   paved  the
possibilities to accelerate  the electrons to GeV  of energies by plasma interaction  with the aforementioned high
power  lasers. Furthermore,  the  acceleration of  the  target  ions to  MeV  of energies  
is  also proved to be feasible  with  existing ultra-intense  lasers.  Depending  on the  laser  and  target 
parameters,  there  are numerous acceleration  mechanisms are  reported, in  line with  the experimental  
findings. The  Target Normal Sheath  Acceleration  (TNSA)  \cite{Snavely2000_PRL,Passoni2010_NJP}, Radiation  
Pressure  Acceleration  (RPA) \cite{Qiao2011_POP,Scullion2017_PRL}, Breakout Afterburner (BOA) \cite{Yin2011_POP,Weng2012_IOP}, 
Relativistic Self Induced Transparency (RSIT) \cite{Siminos2012_PRE,Tushentsov2001_PRL,Juan2017_POP,Choudhary2016_EPJD} 
etc are the name to few.

The high  contrast laser pulses  are desirable for  the studies involving the  interaction with the  thin foil
targets, however, the  prepulse of those high  power lasers is intense  enough to ionize the  target before the
arrival  of the  main pulse \cite{Kaluza2004_PRL}.  The ionization  of the  target  and the  formation of  the plasma  ahead of  the
main  target  has  very  dramatic  consequences  which  in a  sense  can  completely  alter  the  dynamics  of
the  interaction.  The  study  of  the  evolution  of  the  laser  pulse  as  it  propagates  in  the  tenuous
plasma  has drawn  considerable  research interest  around  the globe  both  theoretically and  experimentally
\cite{Davis2005_POP,Sullivan1994_OL,Najmudin2002_TPS,Sprangle2014_POP}. The propagation  of the laser 
pulse  in the under-dense plasma ($n_e  < n_c$) has been  studied in the past \cite{Najmudin2002_TPS,Chen1993_POF}.  The effect 
of the  polarization on the dynamics  of the laser-plasma interaction has been reported in Ref.  \cite{Singh2012_POP}. The influence 
of the magnetic fields on the propagation of the laser in the plasma is discussed in Ref. \cite{Wilson2017_PPCF}. 
The generation of the magnetic fields during intense laser channelling in underdense plasma has been reported in Ref. 
\cite{Smyth2016_POP}. The propagation of the laser or  electromagnetic pulses in plasma also  leads to 
non-linear phenomenon resulting  in the soliton formations \cite{Hadzievski2002_POP,Yin2011_POP}. The existence of the 
solitary waves in the plasma and its effect on the laser pulse itself is  reported in Ref. 
\cite{Sanchez2011_PPCF}. The wakefield generation is  also one of most important phenomenon  as a  consequence 
of the  laser pulse  propagation in the  under-dense plasma \cite{Pukhov2011_PRL}.  As the plasma density approaches the critical 
density $n_c$, the wakefield generation is suppressed and instead laser undergoes nonlinear self-modulation 
\cite{Holkundkar2018_PRE}.

In this work, we study the evolution of  the laser pulse as  it propagates in the underdense  plasma. For the
moderate laser intensities  ($a_0 < 1$) we invoke the  1D relativistic cold fluid model, avoiding  some common approximations relevant for underdense plasma. The results for this case are then compared with the 1D PIC simulations and agreement is  found to be excellent. The evolution of the  ultra-intense ($a_0 > 1$) laser pulse is studied by PIC simulations where the effect of the plasma density, and other laser parameters are also explored. Furthermore, the dispersed ultra-intense laser pulse  is then used to study the acceleration of the ions via relativistic self induced transparency (RSIT).

The organization of our paper is as follows. In Sec. II the governing equations for 1D wave propagation are
discussed along with the details of the PIC simulations. Next, in Sec. III we study the pulse dispersion for
$a_0 < 1$  and $a_0 \gtrsim 10$. The ion  acceleration by the RSIT mechanism via  the dispersed pulses is also 
discussed in  the Sec. III-C followed by the  concluding remarks in Sec. IV.

\section{Theory and simulation model}

\begin{figure*}[t]    \begin{center}   \includegraphics[totalheight=3.5in]{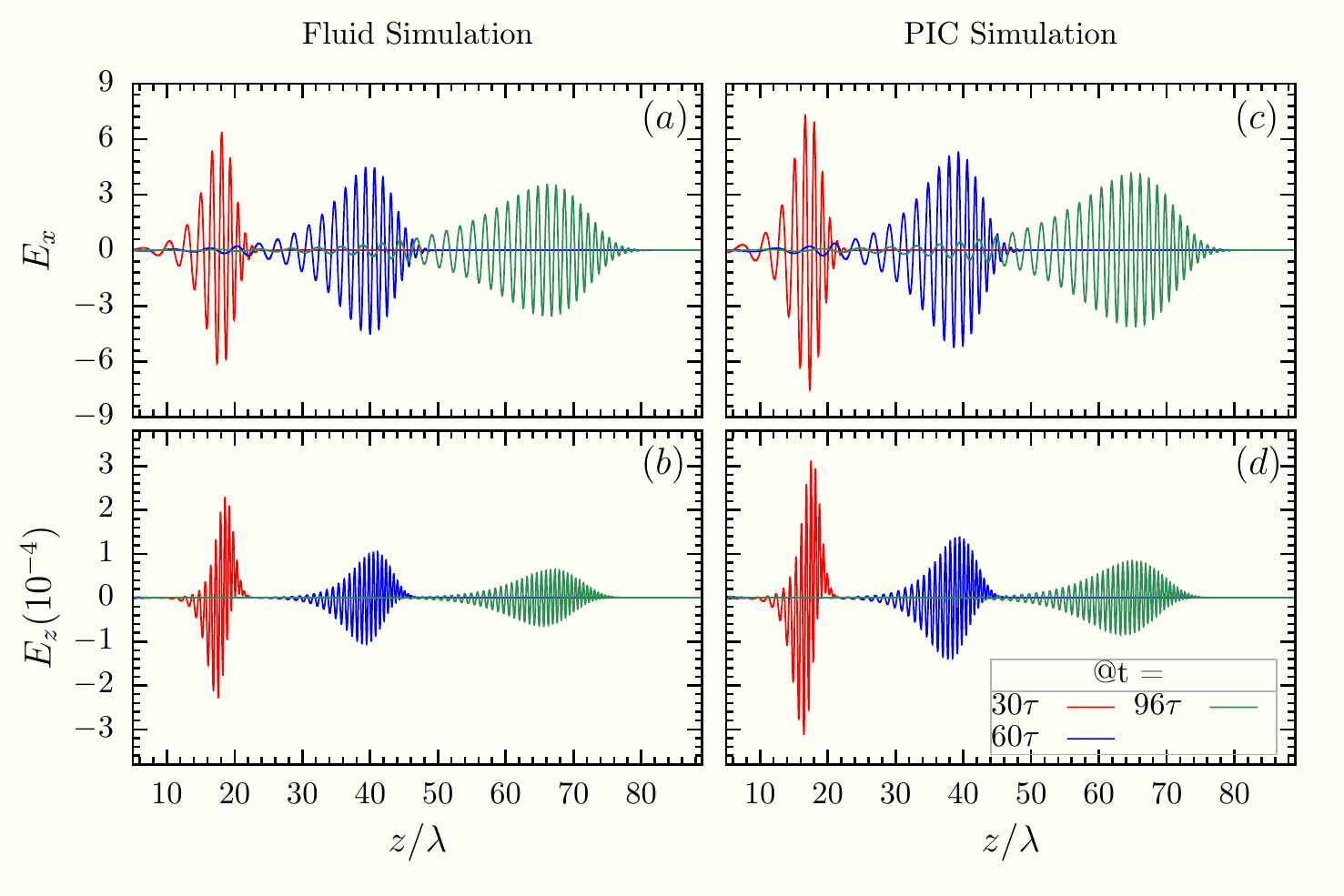}    \end{center}
\vspace*{-7mm} \caption{The spatial profile of the electromagnetic (transverse) fields (a,c) and electrostatic
(longitudinal) fields (b,d) is  presented at different time instances using fluid  simulation (left panel) and
PIC simulation (right panel). Here, we modeled the interaction  of the 800 nm, 3 cycles (FWHM) Gaussian laser
pulse ($a_0 = 0.1$) with plasma having density $0.5n_c$. } \label{dis1} \end{figure*}

\begin{figure}[b]    \begin{center}    \includegraphics[totalheight=2.3in]{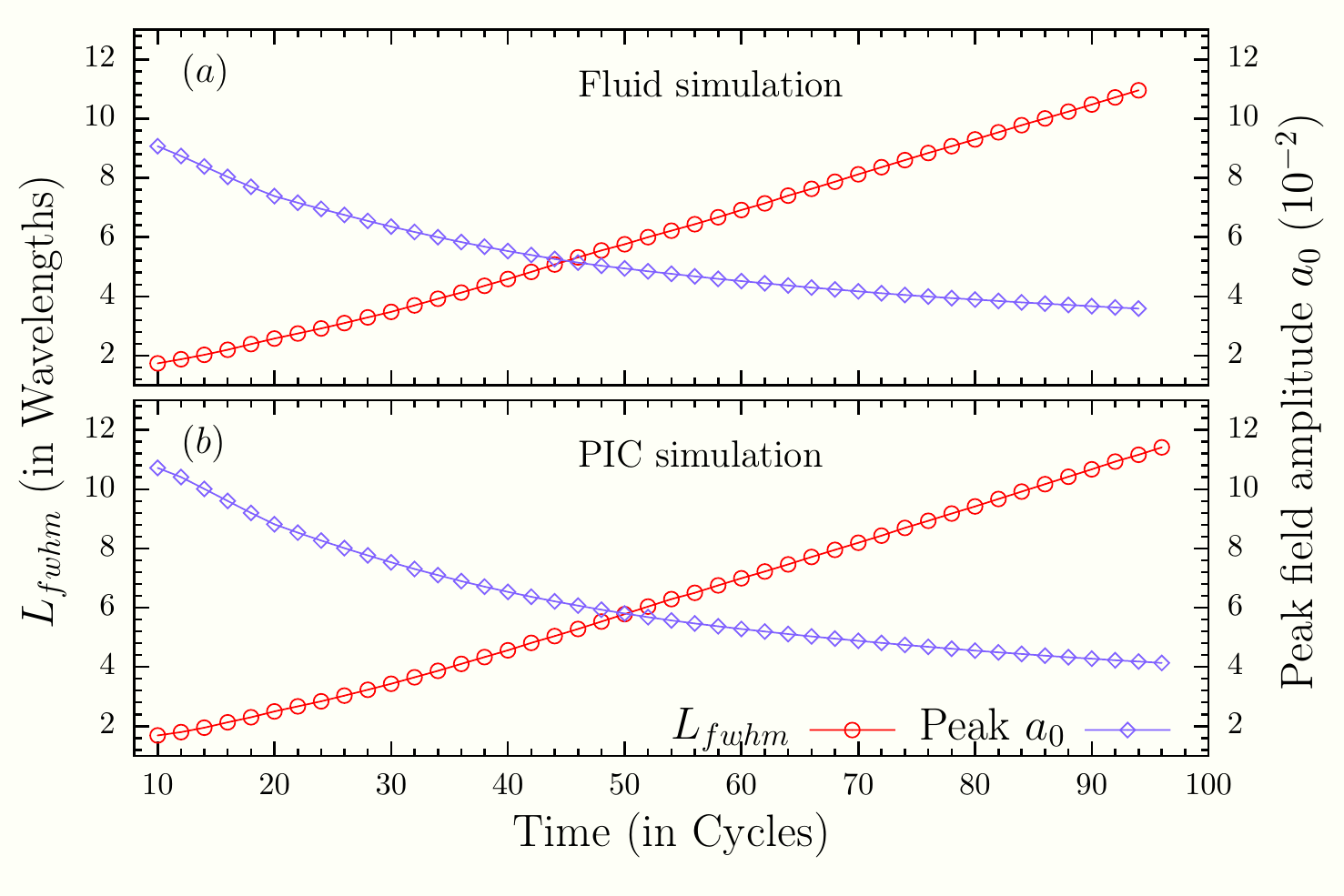}    \end{center}
\vspace*{-7mm} \caption{The temporal evolution of the pulse length  (red circle, left axis) and the peak field
amplitude (blue filled circles, right axis) as it propagates in the underdense plasma is compared by fluid (a)
and PIC simulations  (b). The pulse length $L_{fwhm}$  is estimated in units of the  fundamental wavelength of
the  laser pulse,  similarly time  is presented  in units  of fundamental  laser cycle.  The laser  and plasma
conditions are same as Fig. \ref{dis1}. } \label{dis2} \end{figure}

The objective of this article is to study the dispersion of the electromagnetic (EM) waves as it propagates in
an  under-dense plasma.  The  propagation and  the dispersion  of  the EM  waves  can be  understood by  the
relativistic cold  fluid model.  Recently we  developed a  model to  study the  transition from  the wakefield
generation to the soliton formation \cite{Holkundkar2018_PRE,Brodin2006_IOP}. However, for the sake of 
completeness here also we elaborated  the cold fluid  model. We have  considered the  immobile ions, and  apart 
from this  no further approximations are made. For detailed calculations please refer the \textit{\Amol{Appendix}}.

The laser  amplitude is normalized  as $\mb{a}  = e\mb{A_\perp}/m_ec$,
scalar potential as $\varphi  = e\phi/m_ec^2$, time and space with laser frequency  and wave vector ($\omega t
\rightarrow  t$ and  $k  x \rightarrow  x$)  respectively, velocity  as  $\beta =  \upsilon/  c$, momentum  is
normalized $\mb{p} =  \mb{P}/m_ec$, charge and mass are  normalized by electron charge and  mass, the electron
density is normalized by critical density $n_c = \varepsilon_0 \omega^2 m_e/e^2$. By using these normalization
one can easily deduce from Eq. 30-36 of the \textit{\Amol{Appendix}} the following set of equations 

\be   \frac{\partial^2    \mb{a}}{\partial   z^2}   -    \frac{\partial^2   \mb{a}}{\partial   t^2}    =   n_e
\frac{\mb{a}}{\gamma} \label{sol_0} \ee  
\be \frac{d \beta}{dt} = \frac{(1  - \beta^2)}{\gamma} \frac{\partial
\varphi}{\partial z} - \frac{1}{2 \gamma^2} \Big(  \frac{\partial \mb{a}^2}{\partial z} + \beta \frac{\partial
\mb{a}^2}{\partial t}  \Big) \ee 
\be  \frac{\partial n_e}{\partial  t} + \frac{\partial}{\partial  z} \Big(n_e
\beta\Big) =  0 \ee  
\be \gamma  = \sqrt{\frac{1  + a^2}{1  - \beta^2}  } \label{gg0}\ee  
\be \frac{\partial^2
\varphi}{\partial t\partial z}  = - n_e \beta \label{sol_10} \ee  The above set of equations are  the basis of
our analysis of the dispersion  of the EM wave in the under-dense plasma. In order  to validate the results of
our fluid model,  we used a 1D  particle-in-cell (PIC) simulation, the  details of the PIC  simulations are as
follows: The  1D Particle-In-Cell simulation  (LPIC++) \cite{Lichters1997_MPQ} is  carried out to  compare the
results of the cold fluid model. In this code the electric fields are normalized as we earlier discussed ($a_0
= eE/m_e\omega c$).  However space and time are taken  in units of laser wavelength ($\lambda$)  and one laser
cycle  $\tau  =\lambda /c$  respectively,  mass  and  charge are  normalized  with  electron mass  and  charge
respectively.  We have  used  100 cells  per  laser  wavelength with  each  cell having  50  electron and  ion
macro-particles.  The spatial  grid  size and  temporal  time step  for  the simulation  are  considered to  be
0.01$\lambda$ and 0.01$\tau$ respectively.

\section{Results and Discussions}

We have numerically solved the Eqs (\ref{sol_0})-(\ref{sol_10}) in the same sequence to study the evolution of
the laser pulse entering the simulation box from the  left side. The simulation box of length $100$ $\lambda $
is  considered, with  a constant  unperturbed plasma  density $n_{0}$  throughout the  simulation domain, the
linearly polarized Gaussian laser pulse of wavelength $800$ nm has a full  width half maximum (FWHM) duration
of 3  cycles ($\tau  _{fwhm}=3\times 2\pi $  ). The normalized  amplitude $a_{0}$  is varied in  the different
simulations, and the boundary conditions on  the left side read  as: 
\begin{equation}  \mathbf{a}(0,t)=a_{0}\exp \Big(-\frac{4\log  (2)t^{2}}{\tau
_{fwhm}^{2}}\Big)   \cos   (t)\   \  \mathbf{\hat{x}}   \end{equation}   
\begin{equation}   n_{e}(0,t)=n_{0} \end{equation} 
\begin{equation} \beta(0,t)=\varphi ^{\prime  }(0,t)=0\quad (\varphi ^{\prime }\equiv \partial
\varphi /\partial z). \end{equation} 
It should be noted  that the cold fluid relativistic model is only valid
for the cases when the laser pulse amplitude is $a_0$  is less than unity ($a_0<1$) or for that matter the dispersion is
in the linear regime.  For ultra-intense  laser pulses  $a_0>1$ the  phenomenon of  the wave  breaking and  other
non-linearities limit  the applicability  of the  fluid approach  in describing  the density  modulations. The
results in  this sections are divided  in for the  cases when $a_0<1$ wherein  we compared the results  of the
fluid and PIC  simulations along with the effects of  the laser and plasma parameters on  the dispersion of EM
waves. On the contrary the dispersion of the ultra intense laser pulses ($a_0>1$) is studied by only using PIC
simulations.

\subsection{Pulse dispersion for $a_0 < 1$}

We consider the  propagation of the 800 nm, 3  cycles (FWHM), linearly polarized, Gaussian laser pulse ($a_0 = 0.1$) in the
plasma with an unperturbed plasma  density of $0.5n_c$.  The spatial  profiles of the  transverse EM  fields and
longitudinal electrostatic fields are illustrated at different time instances in Fig. \ref{dis1} both by using
fluid simulation (left panel)  as well as PIC simulations (right panel), and  apparently the agreement between
the two is found to  be good. The dispersive nature of the laser pulse can  be seen by increased pulse length
and decreased  peak amplitudes  as estimated at  different time  instances. As it  propagates deeper  into the
plasma the pulse tend to broaden. Furthermore, it can  be seen from Fig. \ref{dis1}(b) that the wakefield
or longitudinal field generation is suppressed for the chosen laser and plasma parameters and on the contrary
a kind of a localized structure is co-propagating with the laser pulse [see Fig. \ref{dis1}(a)]. The suppression
of the wakefield happens mainly  when the length of the laser pulse ($c\tau_L$)  is larger than the equivalent
length of the plasma oscillations ($\upsilon_g \tau_p$), here $\upsilon_g$ is the group velocity of the plasma
waves  and $\tau_p$  is the  duration of  the one  plasma cycle.  In Ref.  \cite{Holkundkar2018_PRE}, we  have
presented the detailed analysis of the transition from the wakefield to the soliton formation.

For the same laser and plasma parameters ($a_0 = 0.1, \tau_{fwhm} = 3\ \text{cycles}, n_e = 0.5n_c$) the time 
evolution of  the peak field amplitude  and the pulse length  is presented in Fig.  \ref{dis2}, the agreement 
between the fluid and PIC simulations is found to  be excellent. It can be observed from Fig. \ref{dis2}, 
that the pulse length increases almost linearly with time  as the pulse propagates deeper into the plasma, on 
the other  hand, the  peak field  amplitudes decreases.  As we discussed  earlier, for  this laser  and plasma 
parameters the wakefield generation is suppressed and the modulation in plasma density actually co-moves with 
the laser pulse. The total  energy content of the pulse is found to be almost  constant during its passage in 
the plasma, as energy lost to the wakefield generation is almost negligible.                                  

\begin{figure}[t]   \begin{center}   \includegraphics[totalheight=2.3in]{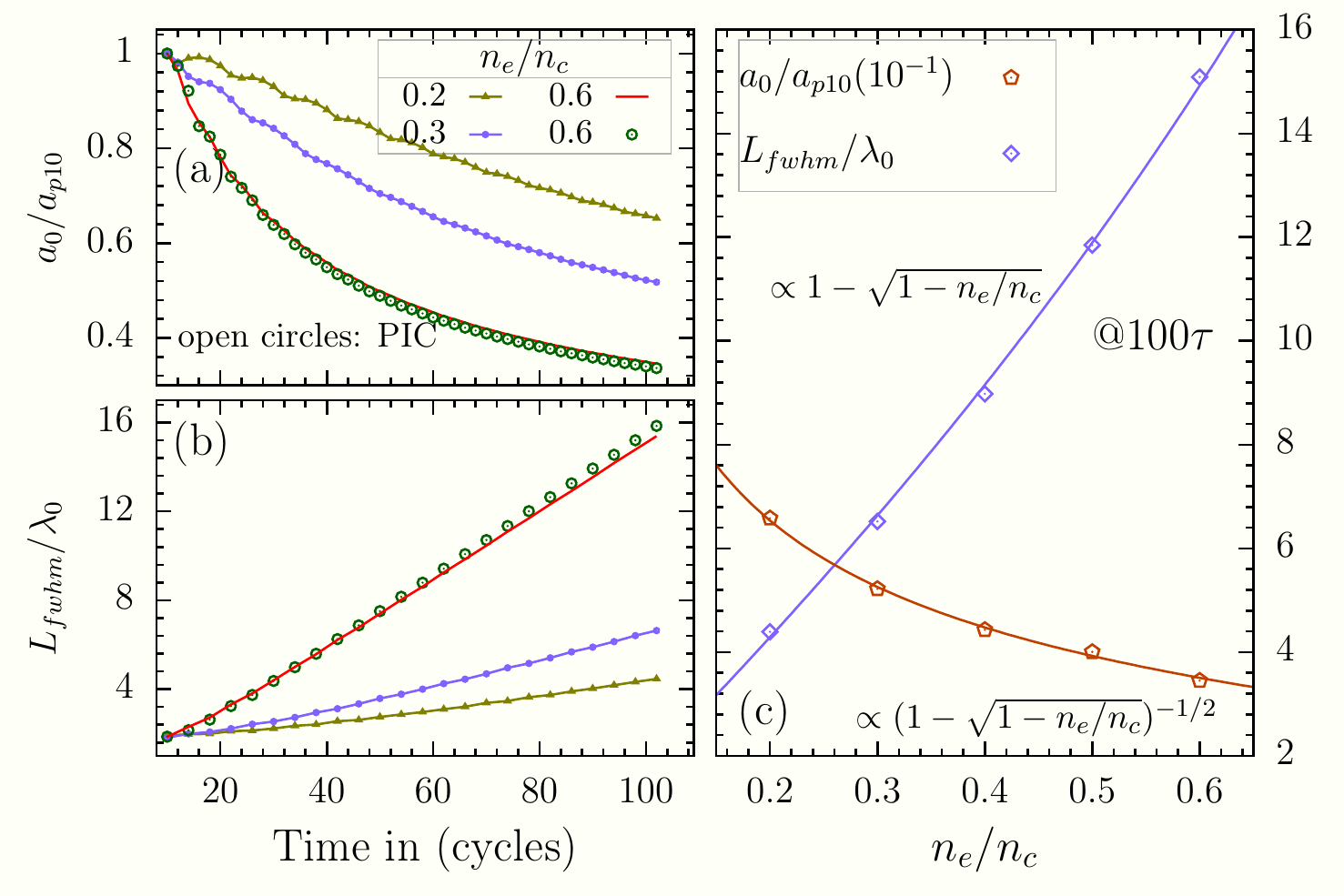}   \end{center}
\vspace*{-7mm} \caption{Temporal evolution of  the peak laser amplitude (a) and pulse  length (b) is presented
for different plasma  densities. The results of the PIC  simulations are also shown with open  circles in both
(a) and (b). The peak amplitude is normalized to the  peak value of the pulse at $t = 10\tau$ ($a_{p10}$). The
value of these  parameters are evaluated at 100$\tau$  are also presented in (c) for  different $n_e/n_c$. The
$a_0 = 0.1$ and $\tau_{fwhm} = 3$ cycles is considered for this case.} \label{den1} \end{figure}

Next, we present the  effect of the plasma density on  the temporal evolution of the pulse length and the
peak field amplitude  of the laser pulse, during its  passage to the uniform density plasma.  For this we have
used the same laser parameters  ($a_0 = 0.1$ and $\tau_{fwhm} = 3$ cycles)  as in Fig. \ref{dis1} and \ref{dis2}.
The peak field amplitude is normalized to its value  at $t=10\tau$ ($a_{p10}$), this has been done to iron out
a slight discrepancy with PIC simulations because anyway here we are more interested in the rate change of the
peak  amplitude as  the pulse  propagates  through the  plasma. The  constant  plasma density  is varied  from
$0.2-0.6n_c$. It can  be observed from Fig. \ref{den1}(b)  that the pulse length increases  linearly with time
and the rate at which  it increases varies with the plasma density. We have compared  the results of our fluid
simulation  with the  PIC simulation  for the  case with  $n_e  = 0.6n_c$  and the  agreement is  found to  be
excellent.  In Fig.  \ref{den1}(c)  we  present the  pulse  length  and the  peak  amplitude  as evaluated  at
$t=100\tau$ for different plasma densities.

\begin{figure}[t]   \begin{center}   \includegraphics[totalheight=2.3in]{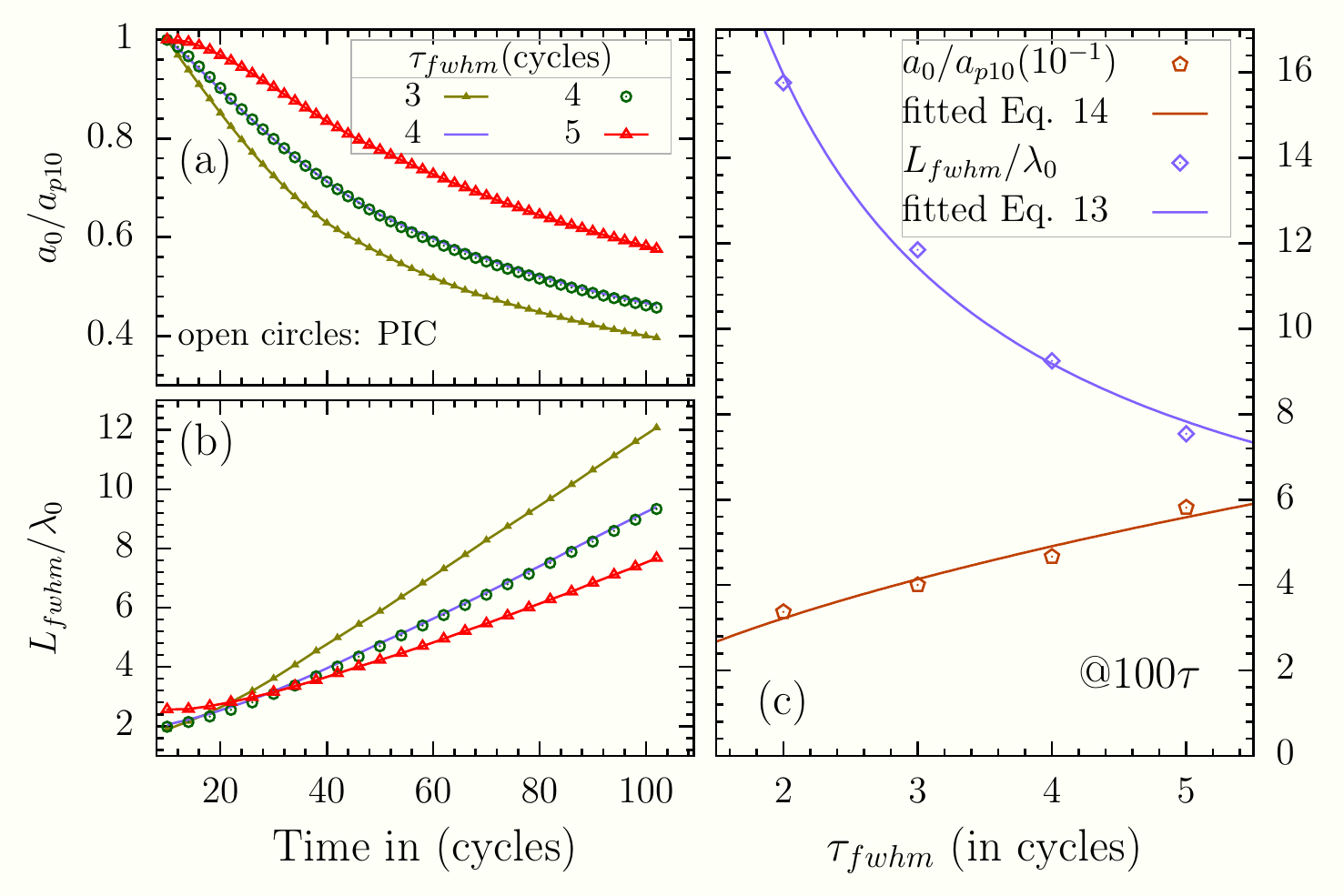}   \end{center}
\vspace*{-7mm} \caption{Temporal evolution of  the peak laser amplitude (a) and pulse  length (b) is presented
for different pulse duration. The results of the PIC simulations are also shown with open circles in both (a)
and (b).  The peak amplitude  is normalized to  the peak value  of the pulse  at it would  be at $t  = 10\tau$
($a_{p10}$). The value of  these parameters are evaluated at 100$\tau$ and are  presented in (c) for different
$\tau_{fwhm}$. The $a_0 = 0.1$ and $n_e = 0.5n_c$ is considered for this case. } \label{tau1} \end{figure}

The linear broadening  of the laser pulse with  time can be understood  in terms of the group  velocity of the
pulse in the plasma. In the linearized theory the group velocity (normalized to $c$) can be calculated as: \be
\upsilon_g = \sqrt{1  - \omega_p^2/\omega^2} = \sqrt{1 -  n_e/n_c}\ee then the time evolution  of pulse length
($L$) can be expected to follow the relation, \be L(t) = L_0 + (1 - \upsilon_g) t \label{length}\ee such, that
in case of vacuum ($\upsilon_g = 1$) the pulse length remains constant, say $L_0$. As we have pointed out that
for this  cases the wakefield generation  is almost suppressed  and the energy  content of the laser  pulse is
almost constant,  this indicates toward  the fact that  as the pulse  broadens, the respective  peak amplitude
should drop accordingly. The energy of the pulse will  scale as $\propto a_0^2 L$, which indicates the drop in
the peak amplitude scales as $\propto 1/\sqrt{L(t)}$.

The scaling is  found to be in accordance  with the results presented in Fig.  \ref{den1}(a). Furthermore, the
variation of  the pulse  length (as evaluated  at $t=100\tau$) with  the plasma  density is presented  in Fig.
\ref{den1}(c). As  expected from Eq. \ref{length},  the pulse length and  pulse amplitude would scale  as: \be
L(n_e) \propto 1 - \sqrt{1 - n_e/n_c} \label{length2}\quad; \quad a_{0} \propto \frac{1}{\sqrt{L(n_e)}}\ee the
fitted Eq. \ref{length2} is also illustrated in Fig. \ref{den1}.

The effect  of the pulse  duration on the   dispersion  is presented in  Fig. \ref{tau1}. Here,  again we
studied the time  evolution of the pulse length and  the peak field amplitude during the  passage of the pulse
through the plasma. We varied  the $\tau_{fwhm}$ for fixed $a_0=0.1$ and $n_e =  0.5n_c$. The results are also
compared with the  PIC simulations as well and an agreement is found to  be excellent. It can be  seen from Fig.
\ref{tau1}(c) that  the length of  the pulse at  say $t =  100\tau$ decreases as  we increase the  laser pulse
duration. It can be understood  as follows, we know the shorter the pulse higher would be the bandwidth,
that translates to the fact that a different portion of  the pulse will propagate with the different velocity and as
a result the larger broadening  of the pulse. On the other hand for longer pulses  the bandwidth is small, and
so the associated dispersion. We can compare the time scales of the laser pulse duration with the time scales
typically involved in the  plasma oscillations $\tau_{fwhm} \propto 1/\omega_p \sim  1/\sqrt{n_e}$ for a rough
estimates related  to the dispersive  nature of the plasma.  \be \tau_{fwhm} \propto  \frac{1}{\sqrt{n_e}} \ee
however, as we saw earlier the pulse length is related  to the plasma density as given by Eq. \ref{length}, so
in  a  sense  $n_e  \propto  L^2$,  this  implies \be  \tau_{fwhm}  \propto  \frac{1}{L}  \implies  L  \propto
\frac{1}{\tau_{fwhm}} \label{len2}\ee and peak amplitude would  be, \be a_0 \propto \frac{1}{\sqrt{L}} \propto
\sqrt{\tau_{fwhm}} \label{taua0}\ee

\begin{figure}[t]   \begin{center}    \includegraphics[totalheight=2.3in]{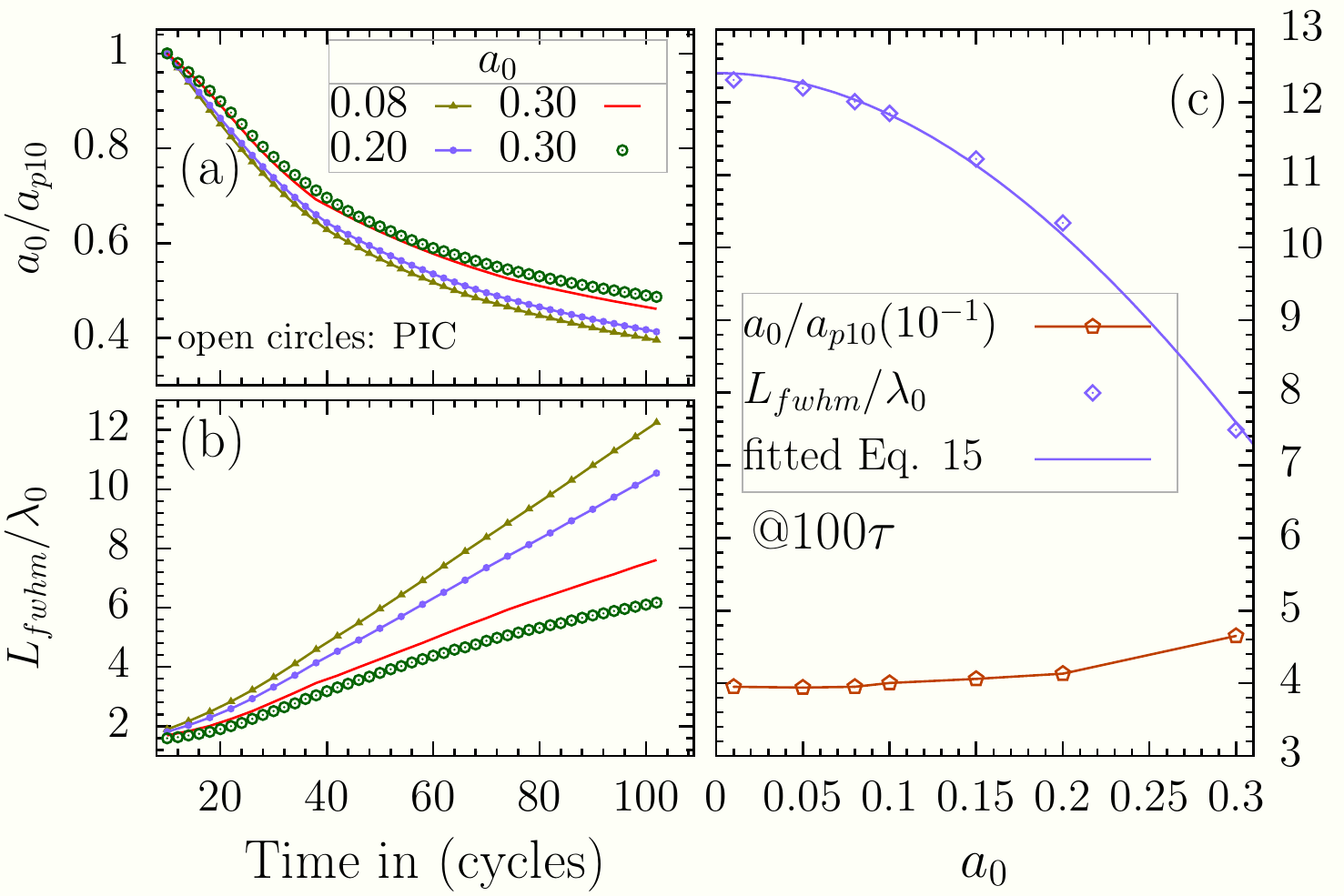}   \end{center}
\vspace*{-7mm} \caption{Temporal evolution of  the peak laser amplitude (a) and pulse  length (b) is presented
for different laser  amplitudes. The results of the PIC  simulations are also shown with open  circles in both
(a) and (b). The  peak amplitude is normalized to the peak  value of the pulse at it would be  at $t = 10\tau$
($a_{p10}$). The value of  these parameters are evaluated at 100$\tau$ and are  presented in (c) for different
$a_0$. The $n_e = 0.5n_c$ and $\tau_{fwhm} = 3$ cycles is considered for this case.} \label{a01} \end{figure}

Next, we  further study  the effect of  the laser  amplitude on the  dispersion of the  laser pulse.  For this
purpose we fixed the pulse duration to $\tau_{fwhm} =  3$ cycles and the plasma density to $0.5n_c$ and varied
the peak  laser amplitude. The time  evolution of the  peak field amplitude and  length of the laser  pulse is
presented in Fig.  \ref{a01}. Again as expected  the linear dispersion law  is found to be  consistent for the
laser and plasma parameters presented. Though for $a_0 =  0.3$ case, we found a bit of discrepancy with fluid
simulation for pulse length evolution, otherwise the rate change of the field amplitude is consistent with the
findings of the PIC simulations.  The value of the field amplitude and pulse  length as evaluated at $100\tau$
is also  illustrated in Fig. \ref{a01}(c).  It is understood that  high intensity laser pulses  tend to disperse
less as  compare to  the low amplitude  pulses, as  a consequence the  pulse length of  the intense  pulses is
smaller than then their low intensity counterpart after  certain time of propagation. As we discussed earlier,
the pulse  length can  be estimated by  Eq. \ref{length},  however with the  relativistic corrections  the Eq.
\ref{length} is modified  as, \be L(t) = L_0  + \Big( 1-\sqrt{1 - \frac{n_e}{\gamma'  n_c}}\Big) t \quad;\quad
\gamma'  \equiv \sqrt{1+a_0^2}  \label{len4}\ee here,  $\gamma'$ is  relativistic factor  (Eq. \ref{gg0}),  we
ignored the  longitudinal motion  of the electrons.  The scaling of  the pulse  length with the  initial laser
amplitude is carried out using Eq. \ref{len4} and the fitted curve is also presented in Fig. \ref{a01}(c).

\subsection{Pulse dispersion for $a_0 > 1$}

In the previous section, we discussed the dispersion of the laser pulses with $a_0 < 1$. We developed an analytical
framework based on the cold relativistic fluid model and benchmarked the results with the 1D PIC simulations. However, for high intense laser pulses, the cold fluid model is no longer valid, as for intense laser fields the nonlinear phenomenon like wave-breaking would prevail, which indeed is outside the purview of the fluid approach. In order to study the dispersion of the intense laser pulses ($a_0 > 1$) we would be using the PIC simulations alone.

\begin{figure}[b]   \begin{center}  \includegraphics[totalheight=2.3in]{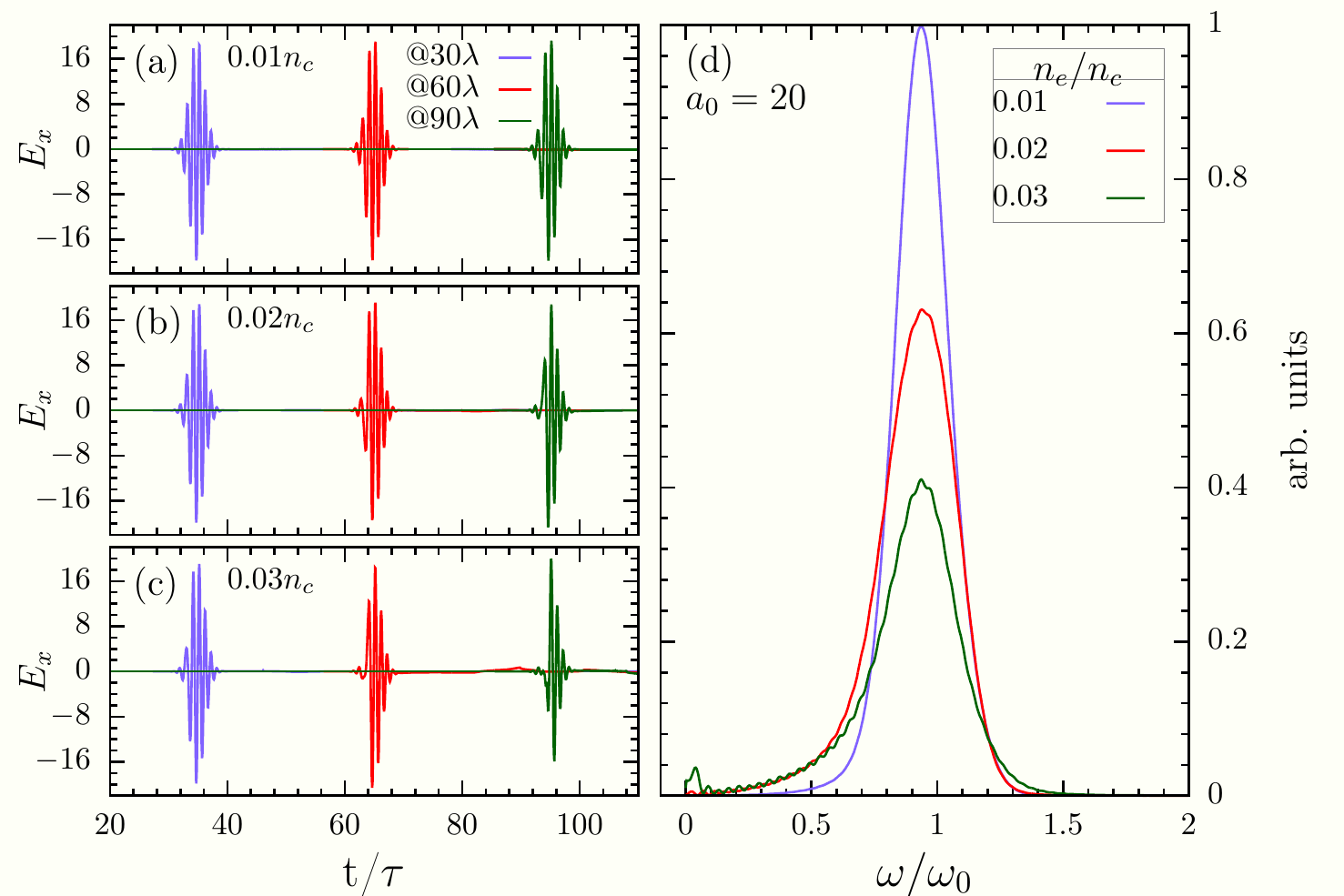}   \end{center}
\vspace*{-7mm} \caption{The temporal  snapshots of the laser field  as evaluated at 30, 60  and 90$\lambda$ is
illustrated  for the  case when  3 cycle  laser with  $a_0 =  20$ is  propagating in  the plasma  with density
0.01$n_c$  (a), 0.02$n_c$(b)  and  0.03$n_c$(c). The  Fourier  spectrum of  the laser  pulse  as evaluated  at
$90\lambda$ is also compared for different plasma densities (d).} \label{Ex-dens} \end{figure}

We consider  the propagation of 3  cycles ($\tau_{fwhm}$), linearly  polarized, Gaussian pulse  with the peak
field amplitudes as $a_0 =  10, 15$ and 20 in the  plasma with uniform density $\lesssim 0.03  n_c$. The reason to
consider the  lower plasma  density (as compared  to the  previous section) for  $a_0 > 1$  is to  mitigate the
formation of the  overdense plasma ($n_e > n_c$)  caused by the ponderomotive force exerted  by intense laser
pulses $a_0  \gtrsim 10$. The overdense  plasma then prohibits the further propagation of the laser pulses,
till it becomes sufficiently underdense (by space-charge effect)  to allow the passage of the laser pulse. For
this kind of scenario we might have the reflections of  the laser pulse from the different part of the plasma,
where it turns overdense. In order to avoid any reflections by the formation of the overdense plasma, in this
section we would be considering the plasma densities $\lesssim 0.03 n_c$.

  \begin{figure}[t] \begin{center}  \includegraphics[totalheight=2.3in]{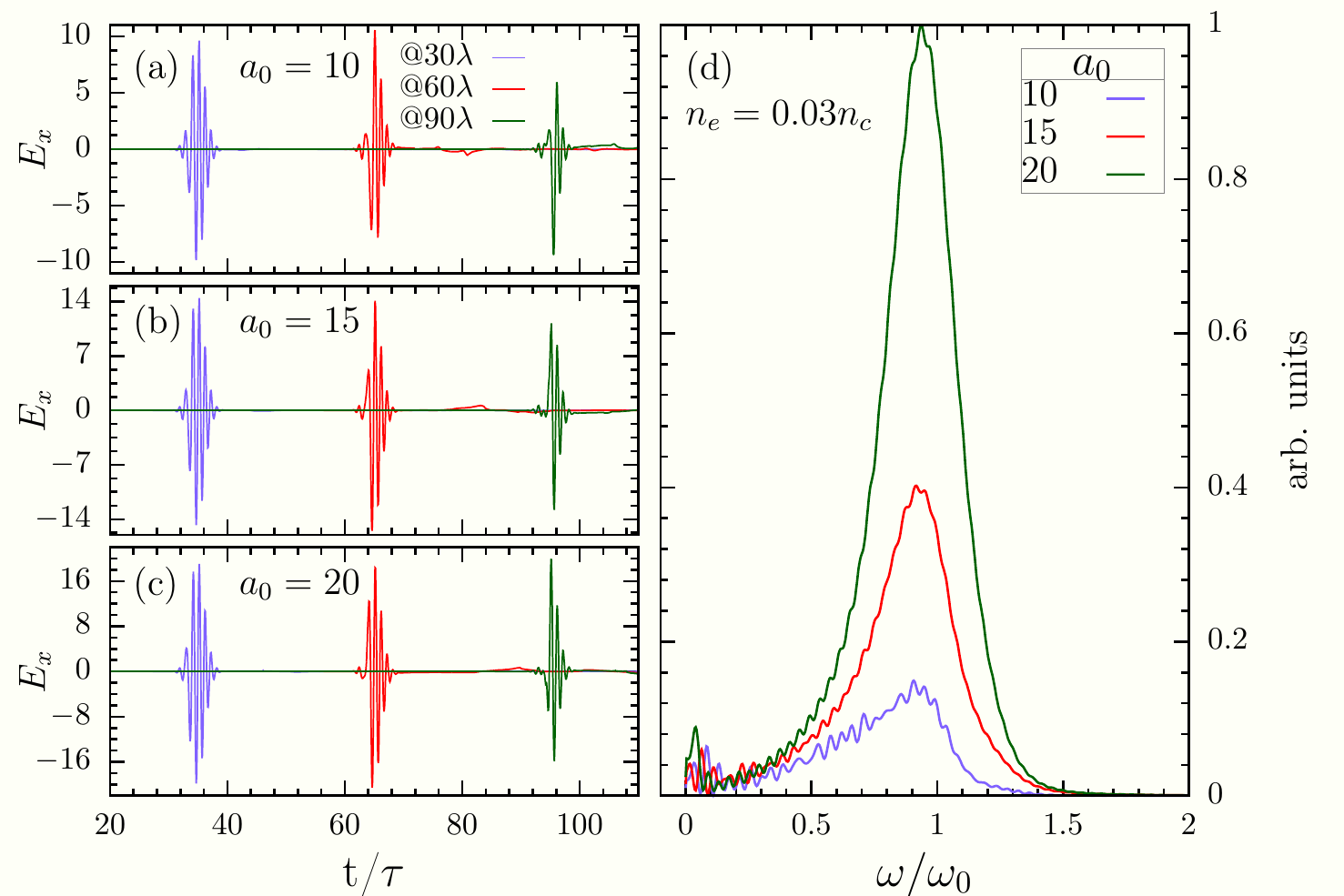} \end{center}
\vspace*{-7mm} \caption{The temporal  snapshots of the laser field  as evaluated at 30, 60  and 90$\lambda$ is
illustrated for  the 3 cycle laser propagating in the  plasma with density 0.03$n_c$.  The laser amplitude
$a_0 =  10$ (a),  15(b) and  20(c) are  considered. The Fourier  spectrum of  the laser  pulse as  evaluated at
$90\lambda$ is also compared for different laser amplitudes (d).} \label{Ex-a0} \end{figure}

We compared the time evolution  of the laser ($a_0 = 20$) electric field  for three different plasma densities
in Fig. \ref{Ex-dens}. The field profiles are  evaluated after the laser propagated the distances $30\lambda$,
$60\lambda$ and $90\lambda$ in the  plasma. It can be observed from this figure that  the peak of the envelope
moves roughly with the  same velocity  for different  plasma densities,  this indicates  that the  group
velocity of the laser is  more or less unaffected for the considered laser and  plasma parameters, or maybe it
would require longer simulation to see any prominent effect on propagation. We have also presented the Fourier
spectrum of the laser pulse in Fig. \ref{Ex-dens}(d). It  can be seen that for higher densities the broadening
of the frequency spectrum is larger because of the stronger plasma wave generation.  The spectrum is found to
be red shifted in direct correlation with the plasma  density \cite{Pathak2018_POP}. 
The red shifting and broadening of the spectrum generally accounts  for the stronger plasma wave generation, 
because of the energy transformation from the laser to the plasma. This fact can also be observed from the 
Fig. \ref{Ex-dens}(d), wherein the red shift for higher density plasma is larger as compared to the lower
density plasma. In order to elucidate the effect of the laser pulse amplitude on the dispersion of the laser
pulse, in Fig. \ref{Ex-a0} we have varied the laser pulse amplitude while keeping  the plasma density fixed
at $0.03n_c$. The broadening of the spectrum is seen to be prominent for the $a_0  = 10$ as compare to $a_0 =  20$,
because the rate at which the  energy is depleted for lower laser amplitudes would be larger as compared to higher laser amplitudes.

\begin{figure}[b]  \begin{center}   \includegraphics[totalheight=2.5in]{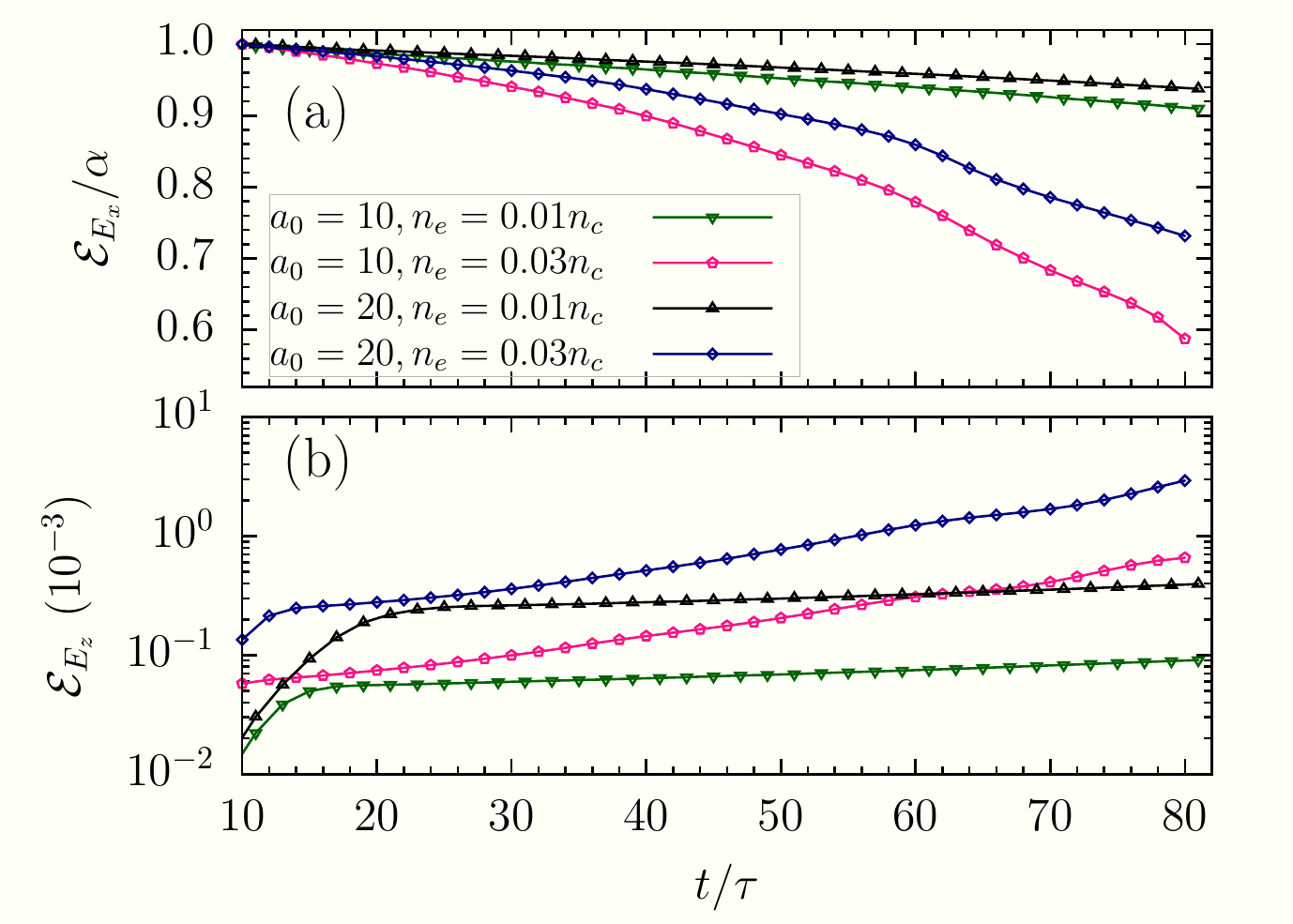}  \end{center}
\vspace*{-7mm} \caption{Temporal evolution of the laser pulse  energy (a) and the longitudinal field energy is
presented for  3 cycle  Gaussian pulse with  $a_0 =  10,\ 20$ when  it propagated in  the plasma  with density
$0.01,\ 0.03 n_c$. The laser pulse energy in (a) is normalized to maximum value at $t = 10\tau$, as we 
are interested in the depletion rate of the laser pulse energy for different laser and plasma parameters.}  
\label{pulse-ene} \end{figure}

The   time  evolution   of  the   electromagnetic   and  electrostatic   field  energies   are  presented   in
Fig.\ref{pulse-ene}. Here, again we  have considered the propagation of the 3 cycle, linearly polarized laser
pulse with $a_0 = 10,20$  in the plasma having densities $0.01,0.03n_c$. As time  progress the decrease in the
electromagnetic field energy and increase in the electrostatic field energy is observed which indicates toward
the stronger  plasma wave  generation at the  cost of the  electromagnetic energy.  As expected it is further
observed that the depletion rate of the electromagnetic field energy is larger  for  the laser with peak amplitude
$a_0 = 10$ as compare to $a_0 = 20$. This is so because the dispersion of the high intensity laser pulses would
be relatively slower than the laser pulses with lower intensity. The direct correlation of  the plasma
density can also be seen on the depletion rate of the electromagnetic field energy, and so the growth in the
longitudinal field energy.

\subsection{Ion acceleration by intense dispersed pulses}

\begin{figure}[t]  \begin{center}  \includegraphics[totalheight=2.3in]{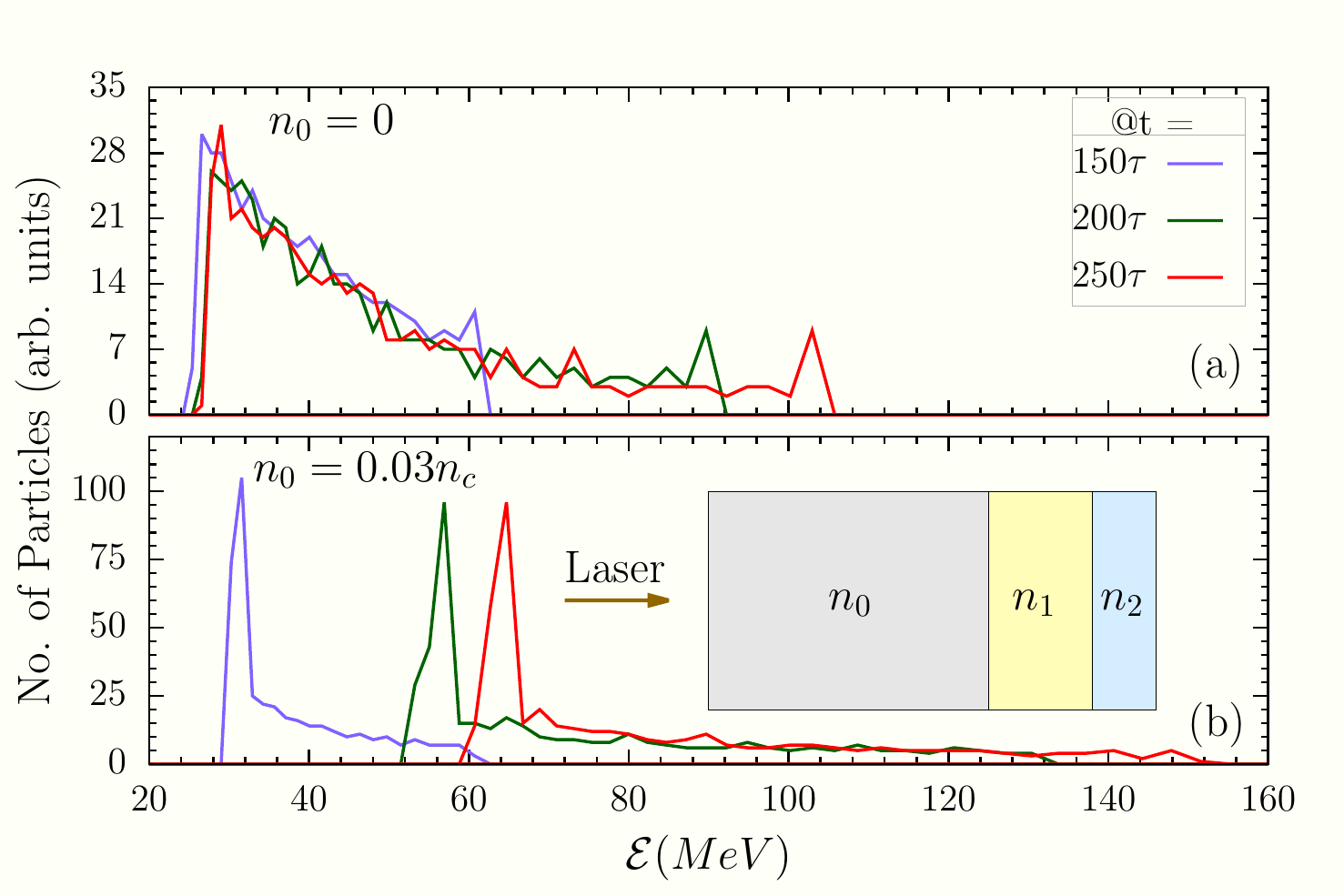}  \end{center}
\vspace*{-7mm} \caption{The energy spectrum  of the ions from the secondary  layer is
compared when the laser directly interacts with the target i.e. $n_0 = 0$ (a) and when initially its 
allowed to propagate in the 100$\lambda$ pre-plasma ($n_0 = 0.03n_c$) prior to interaction with composite target (b). 
The geometry of the setup is also illustrated as an inlet. Here, $n_0, n_1$ and $n_2$ are the plasma 
density of the pre-plasma, first layer and second layer respectively. Please refer the text for the 
physical parameters of first and second layer.} 
\label{enersec} 
\end{figure}

We have recently  demonstrated the use of the negatively  chirped laser pulses to accelerate the  ions to a few 
hundreds of the  MeV by using a  double layer (Hydrogen plasma) target  \cite{Choudhary2018_POP}. The primary 
layer having density  $6n_c$ is found to  be transparent for the  negatively chirped laser pulse  with $a_0 = 
20$, creating a persistent electrostatic field  which actually accelerates the  ions from the  secondary layer 
($0.1n_c$). Next, we deploy the  similar geometry of two layer target just after  the low density plasma. The 
propagation of the laser pulse in underdense plasma actually  causes the dispersion of the pulse, as a result 
the pulse would  be chirped when it incidents on the two-layer target. In Fig. \ref{enersec} we present the 
energy spectrum of  the ions from the  secondary layer. We considered  the 3 cycle Gaussian laser pulse with 
$a_0 =  20$ propagates in the  100$\lambda$ long underdense  plasma having density $0.03n_c$.  The dispersed 
pulse then  incidents on the two-layer target, first  layer is $0.75\lambda$ thick with density $6n_c$ and 
adjacent secondary layer is $0.2\lambda$ thick with density $0.1n_c$.

The  energy spectrum  of the  ions from  the secondary  layer  is compared  for the  cases when  there is  no 
underdense plasma and when the laser propagated through the plasma [see, Fig. \ref{enersec}]. It can be seen 
that the ions  from the secondary layer  are very efficiently accelerated to almost mono-energetically when 
the dispersed pulse interacts  with the two-layer composite target geometry. The reason being the dispersion, 
wherein the  frequency of  the pulse undergone  the modulation  in space  and time, in  other words  pulse is 
somewhat chirped. As the high frequency component of the pulse interacts with the primary layer, it transmits 
through  the layer  by the  relativistic self induced transparency, or in other words, the critical density for 
the transmission gets modified for the chirped laser pulses. The  transmitted pulse drags the electrons from  
the primary (as well as secondary) layer with them, creating very persistent longitudinal electrostatic field \cite{Choudhary2018_POP}. 
The electrostatic field then pulls the ions  from the secondary  layer, forming the  mono-energetic ion  bunch.
However, in the absence of the pre-plasma the primary target is opaque to the incident unchirped pulse, resulting 
in the reflection. If the laser pulse suffers the reflection at the primary layer then acceleration is mostly 
caused by the radiation pressure mechanism, resulting in lower energy yield for the same laser intensity 
[see, Fig. \ref{enersec}(a)]. The optimization of the degree of the pulse chirping (dispersion) by varying the 
pre-plasma length and/or density for most efficient acceleration of the ions from the secondary layer is beyond 
the scope of the current manuscript.

\section{Concluding Remarks}

We have  studied the  dispersion of  the laser  pulse as  it propagates  in the  underdense plasma.  For the 
moderate laser intensities ($a_0 <  1$) we invoked a 1D relativistic cold fluid model to evaluate the spatial 
and temporal evolution  of the laser as  it propagates in the  plasma with density $\lesssim  0.6 n_c$. Apart 
from the immobile ions, no further approximations are made. The effect of the laser pulse amplitude, pulse 
duration and the  plasma density is explored using the fluid model and the results are  compared with the 1D 
PIC  simulations along  with  the expected scaling  laws. The agreement between fluid  model  and the  PIC 
simulations are found to be excellent. Furthermore, in order to study the interaction of highly intense laser 
pulses $a_0 \gtrsim  10$, we only relied  on the PIC simulations  as the nonlinear nature  of the interaction 
process is beyond  the validity of the  cold relativistic fluid model.  For these cases we  restricted to the 
plasma density  $\lesssim 0.03n_c$, or  the strong ponderomotive  force of laser  pulses tend to  make plasma 
over-dense ($n_e  > n_c$) restricting  the further propagation  of the laser  pulse. The conversion  from the 
electromagnetic  field energy  to  the electrostatic  fields  in the  form  of plasma  waves  results in  the 
dispersion and so the red shift of the pump laser pulses. The dispersed pulse then allowed to be incident on the 
sub-wavelength two layer composite target. The ions from the  thin, low density secondary layer are found to be 
mono-energetically accelerated to $\sim 70$ MeV, which was not the case without the dispersion.

\section*{Appendix-I}

Maxwell's equation using Coulomb gauge can be written as [J D Jackson, \emph{Classical Electrodynamics}],
\be \nabla^2 \phi = -\rho / \varepsilon_0\ee 
\be \nabla^2 \mb{A} - \frac{1}{c^2} \frac{\partial^2 \mb{A}}{\partial t^2} 
   = -\mu_0 \mb{J} + \frac{1}{c^2} \nabla \Big( \frac{\partial \phi}{\partial t}\Big)\ee 
   
We are considering the 1D case so that the variation of $\phi$ and $\mb{A}$ along $x$ and $y$ are not 
considered. The electron current density can be written as $\mb{J} = -e n_e \mb{v}$. Furthermore, $\mb{v} = 
\mb{v}_{||} + \mb{v}_\perp$ and $\mb{v_z} \equiv \mb{v}_{||}$. By using these approximations above equations 
can be written as,

\be  \frac{\partial^2 \phi}{\partial z^2}   = -\rho / \varepsilon_0\ee 
\be \frac{\partial^2  \mb{A}}{\partial z^2}  - \frac{1}{c^2} \frac{\partial^2 \mb{A}}{\partial t^2} 
   = \mu_0 e n_e (\mb{v}_z + \mb{v}_\perp) + \frac{1}{c^2} \frac{\partial}{\partial z} \Big( \frac{\partial 
\phi}{\partial t}\Big)\ee

The last equation can be split into two parts, one for perpendicular and another for parallel directions 
respectively. It should be noted that here parallel and perpendicular directions are taken with respect to 
the direction of laser pulse propagation. A Laser is polarized in a plane perpendicular to $z$ axis and hence 
it will contribute toward the perpendicular component of the particle velocities  ($\mb{v}_\perp$). On the 
other hand the electrostatic potential created would be responsible for the motion of the particle along the 
$z$ direction. By equating the perpendicular and parallel component from RHS and LHS one obtains,
\be \frac{\partial^2  \mb{A}}{\partial z^2}  - \frac{1}{c^2} \frac{\partial^2 \mb{A}}{\partial t^2} 
   = \mu_0 e n_e \mb{v}_\perp \ee
\be \frac{1}{c^2} \frac{\partial^2 \phi}{\partial t\partial z} = - \mu_0 e n_e \upsilon_z  \ee    
here we have used $\mb{v_z} = \upsilon_z \mb{e_z}$ 

A Laser is considered to be propagating along the $z$ direction and hence denoted by,
\be \mb{A} = A_x \mb{e_x} + A_y \mb{e_y}\ee
here $\mb{e_x}$ and $\mb{e_y}$ are unit vectors along $x$ and $y$ directions respectively. The electric and 
magnetic fields are denoted by,
$ \mb{E} = -\nabla \phi - \partial \mb{A}/\partial t$
, and $ \mb{B} = \nabla \times \mb{A}$. The electric field can further be written as, $\mb{E}_{||} = -\nabla 
\phi$ (Wakefield) and 
$\mb{E}_{\perp} = - \partial \mb{A} /\partial t $ (Laser electric field).

Now consider the Lorentz force equation,
\be \frac{d\mb{P}}{dt} = -e \Big[-\nabla \phi - \frac{\partial \mb{A}}{\partial t} 
                        + \mb{v\times (\nabla \times \mb{A} ) }\Big] \ee
It should be noted that $\mb{A}$ only varies along $z$ direction and only contains the perpendicular 
components and hence $\partial_x = \partial_y = A_z = 0$.
The above equation can be written as,
\be
\mb{v\times (\nabla \times A)} = -\upsilon_z \frac{\partial \mb{A_\perp}}{\partial z} 
                                 + \Big(\mb{v_\perp} \cdot \frac{\partial \mb{A_\perp}}{\partial z}\Big)  
\mb{e_z}
\ee  
and hence, the perpendicular and parallel component of the Lorentz force equation respectively can be written as 
\be \frac{d \mb{P_\perp}}{dt} = e \Big[ \frac{\partial }{\partial t} + \upsilon_z \frac{\partial }{\partial 
z} \Big]  \mb{A_\perp} = \frac{d}{dt} (e\mb{A_\perp} )
\label{pperp}\ee
\be \frac{d \mb{P_z}}{dt}  = e  \frac{\partial \phi}{\partial z}\mb{e_z}    - e  \Big(\mb{v_\perp} \cdot 
\frac{\partial \mb{A_\perp}}{\partial z}\Big) \mb{e_z}  \label{pz}\ee
which translates to the fact that,
\be \mb{P_\perp} = e \mb{A_\perp} \implies \mb{v_\perp} = e\mb{A_\perp}/\gamma m_e\ee 
Substituting the value of $\mb{v_\perp}$ from (\ref{pperp}) to (\ref{pz}) one obtains (we omit the $\mb{e_z}$ 
for the 
sake of convenience, as all the quantities are along $z$ direction only),
\be \frac{d P_z}{dt} = e  \frac{\partial \phi}{\partial z}   
    - \frac{e^2}{2 \gamma m_e}  \frac{\partial \mb{A}_\perp^2}{\partial z}    
\label{pz2}
\ee
The last term in (\ref{pz2}) is the Ponderomotive force which is responsible for the displacement of the 
electrons
from the laser focus.  

We need to solve for the $\upsilon_z$ so in view of this (\ref{pz2}) can further be written as,
  
\be \frac{d P_z}{dt} = \frac{d}{dt} (\gamma m_e \upsilon_z)   
    = e  \frac{\partial \phi}{\partial z}  
    - \frac{e^2}{2 \gamma m_e}  \frac{\partial \mb{A}_\perp^2}{\partial z}     
\ee

\be \gamma \frac{d \upsilon_z}{dt}  + \upsilon_z \frac{d \gamma}{dt} 
    = \frac{e}{m_e}  \frac{\partial \phi}{\partial z}  
    - \frac{e^2}{2 \gamma m_e^2}  \frac{\partial \mb{A}_\perp^2}{\partial z}     
\ee

\be \frac{d \upsilon_z}{dt}  
    = \frac{e}{\gamma m_e}  \frac{\partial \phi}{\partial z}  
    - \frac{e^2}{2 \gamma^2 m_e^2}  \frac{\partial \mb{A}_\perp^2}{\partial z}
    -  \frac{\upsilon_z}{\gamma} \frac{d \gamma}{dt}      
\label{vz}
\ee
The rate of change of total energy ($\mathcal{E} = \gamma m c^2$ ) of the charge particle can be written as,
\be \frac{d\mathcal{E}}{dt} = q \mb{v}\cdot \mb{E} \ee
which in our case ($q = -e$, $m = m_e$ and $\mb{E} = -\nabla \phi - \partial \mb{A}/\partial t $) can be 
written as,
\be \frac{d\gamma}{dt} = \frac{-e}{m_e c^2} (\mb{v_\perp} + \mb{v_z}) \cdot ( -\nabla \phi - \partial 
\mb{A}/\partial t)\ee

\be \frac{d\gamma}{dt} = \frac{e}{m_e c^2} \Big( \upsilon_z \frac{\partial \phi}{\partial z} 
  + \mb{v_\perp} \cdot \frac{\partial \mb{A}_\perp}{\partial t} \Big)
\ee
Using again the value of $\mb{v_\perp}$ from (\ref{pperp}) in above equation simplifies to,

\be \frac{d\gamma}{dt} = \frac{e}{m_e c^2} \Big( \upsilon_z \frac{\partial \phi}{\partial z} 
  + \frac{e}{2\gamma m_e}\frac{\partial \mb{A}^2_\perp}{\partial t} \Big)
\label{gama0}  
\ee

Substituting (\ref{gama0}) in (\ref{vz}) one obtains,

\be \frac{d \upsilon_z}{dt}  
    = \frac{e}{\gamma m_e}  \Big(1 - \frac{\upsilon_z^2}{c^2}\Big) \frac{\partial \phi}{\partial z}  
    - \frac{e^2}{2 \gamma^2 m_e^2}  \Big( \frac{\partial \mb{A}_\perp^2}{\partial z} 
    + \frac{\upsilon_z}{c^2}\frac{\partial \mb{A}_\perp^2}{\partial t} \Big) 
\ee

 Local charge separation results in electrostatic fields which can be taken into account by Poisson's 
equation,
\be
 \nabla^2 \phi = - \frac{e}{\varepsilon_0} (Z n_i - n_e)
\ee
here $Z$ and $n_i$ are ion charge and density respectively, however, $n_e$ is electron density. In 1D case 
$\nabla^2$ 
is replaced by $\partial^2/\partial z^2$.

\be
 \frac{\partial^2\phi}{\partial z^2}  = - \frac{e}{\varepsilon_0} (Z n_i - n_e)
\ee

 The charge is conserved by continuity equation,
\be
\frac{\partial n_e}{\partial t} + \nabla \cdot (n_e \mb{v}) = 0
\ee 
which again for 1D case can be written as,

\be
\frac{\partial n_e}{\partial t} + \frac{\partial}{\partial z} (n_e \upsilon_z) = 0
\ee 

Now we deduce the expression for $\gamma$ in terms of vector potential,
By definition,
\be \gamma = \frac{1}{\sqrt{1 - (\upsilon_\perp^2 + \upsilon_z^2)/c^2}}\ee 
\be \upsilon_\perp^2 + \upsilon_z^2  = c^2 (1 - 1/\gamma^2)\ee
Using the fact that $\mb{v}_\perp = e \mb{A}_\perp/\gamma m_e$ we obtain,
\be \frac{e^2A_\perp^2}{\gamma^2 m_e^2} + \upsilon_z^2  = c^2 \Big(1 - \frac{1}{\gamma^2} \Big)\ee
Solving for $\gamma$ one obtains,

\be \gamma = \sqrt{\frac{1 + (eA_\perp/m_ec)^2}{1 - \upsilon_z^2/c^2} }\ee

So finally the complete set of equations can be summarized as follows,  
\be \frac{\partial^2  \mb{A}_\perp}{\partial z^2}  - \frac{1}{c^2} \frac{\partial^2 \mb{A}_\perp}{\partial 
t^2} 
   = \mu_0 e n_e \mb{v}_\perp \label{aa0}\ee
\be \frac{1}{c^2} \frac{\partial^2 \phi}{\partial t\partial z} = - \mu_0 e n_e \upsilon_z  \ee  

\be \mb{P_\perp} = e \mb{A_\perp} \implies \mb{v_\perp} = e\mb{A_\perp}/\gamma m_e \ee

\be \frac{d \upsilon_z}{dt}  
    = \frac{e}{\gamma m_e}  \Big(1 - \frac{\upsilon_z^2}{c^2}\Big) \frac{\partial \phi}{\partial z}  
    - \frac{e^2}{2 \gamma^2 m_e^2}  \Big( \frac{\partial \mb{A}_\perp^2}{\partial z} 
    + \frac{\upsilon_z}{c^2}\frac{\partial \mb{A}_\perp^2}{\partial t} \Big) 
\ee

\be
 \frac{\partial^2\phi}{\partial z^2}  = - \frac{e}{\varepsilon_0} (Z n_i - n_e)
\ee

\be
\frac{\partial n_e}{\partial t} + \frac{\partial}{\partial z} (n_e \upsilon_z) = 0
\ee 

\be \gamma = \sqrt{\frac{1 + (eA_\perp/m_ec)^2}{1 - \upsilon_z^2/c^2} } \label{gg0f}\ee

These are the complete set of equations in closed form which need to be solved numerically 
with appropriate boundary conditions.

\section*{Acknowledgments} Authors  would like to  acknowledge the Department  of Physics, Birla  Institute of
Technology  and  Science,  Pilani,  Rajasthan,  India  for the  computational  support.  AH  acknowledges  the
computational resources funded by the DST-SERB project EMR/2016/002675.


%

\end{document}